\documentclass[aps,prl,floats,twocolumn,showpacs,superscriptaddress]{revtex4-2}

\usepackage{graphics}
\usepackage{amsmath}
\usepackage{amsfonts}
\usepackage{amssymb}
\usepackage{times}
\usepackage{graphicx}% Include figure files
\usepackage{dcolumn}% Align table columns on decimal point
\usepackage{bm}% bold math
\usepackage{color}
\usepackage{comment} %Quitar cuando toque
\usepackage{subcaption}
\begin{document}

%\preprint{APS/123-QED}

\title{Directionality-induced jamming in multiplex networks}% Force line breaks with \\

\author{Mateo Bouchet}
\author{Alejandro Tejedor}%
\affiliation{%
 Institute for Biocomputation and Physics of Complex Systems (BIFI), Universidad de Zaragoza, 50018 Zaragoza, Spain\\
}%
\affiliation{
  Department of Theoretical Physics, University of Zaragoza, 50009 Zaragoza, Spain
}

\author{Xiangrong Wang}%
\affiliation{
 College of Mechatronics and Control Engineering, Shenzhen University, Shenzhen 518060, China\\
}

\author{Yamir Moreno}%
\affiliation{%
 Institute for Biocomputation and Physics of Complex Systems (BIFI), Universidad de Zaragoza, 50018 Zaragoza, Spain\\
}%
\affiliation{
  Department of Theoretical Physics, University of Zaragoza, 50009 Zaragoza, Spain
}%

\date{\today}% It is always \today, today,

\begin{abstract}
We study diffusion on multiplex networks with directed interlayer couplings. We demonstrate both numerically and analytically that even with undirected layers, interlayer directionality alone reproduces superdiffusion and the prime regime. We further reveal a new phenomenon, the directionality-induced jamming, whereby directed interlayer links hinder diffusion, fragmenting the system into dynamically disconnected components and preventing convergence to the steady state of the diffusion process. Via an optimization process, we show that this new regime is attainable in both toy models and real-world topologies. These findings underscore the crucial role of interlayer link directionality in shaping the emergent behavior of multiplex systems, with potential implications for the design and control of such systems.
\end{abstract}

\maketitle

Complex systems often exhibit multiple types of interactions operating simultaneously, making network theory \cite{Watts1998,Barabasi1999,Newman2010,Boccaletti2006,Barrat2008,Rodriguez-Iturbe1997,Bullmore2009, Tejedor2017}, and in particular, the framework of multilayer and multiplex networks \cite{Boccaletti2014, Bianconi2018, AletaMoreno2019, DeDomenico2023,Domenico2013, Kivela2014, AletaMoreno2019}, an essential tool to describe their structure and dynamics. In a multiplex representation, the same set of nodes is connected through distinct layers, each encoding a different interaction pattern. This architecture captures the intertwined nature of real systems such as social \citep{Cozzo2013,Li2015,Arruda2017, Smith-Aguilar2019}, transportation \citep{Aleta2017, Tejedor2018GRL}, and biochemical networks \citep{Cozzo2012,Battiston2017, Valdeolivas2018}, where interdependencies between layers give rise to collective behaviors unattainable in single-layer topologies.

Diffusion is among the most fundamental dynamical processes studied in this context. Despite its apparent simplicity, diffusion on multiplex networks reveals a rich phenomenology driven by the interplay between topology and coupling strength. Gómez et al. \citep{Gomez2013} showed that interlayer coupling can enhance transport efficiency, leading to {\it superdiffusion} $-$a regime where diffusion in the multiplex outpaces that in any individual layer. Later, Tejedor et al. \citep{Tejedor2018} demonstrated that when directionality within layers is introduced, diffusion can become non-monotonic with coupling, producing the {\it prime regime}, where intermediate coupling yields the fastest relaxation (see also \cite{Wang2021}). These results unveiled how structural asymmetries control diffusion speed, yet the role of directionality in {\it interlayer} couplings has remained largely unexplored.

In this work, we address this knowledge gap by investigating diffusion on multiplex networks where interlayer links themselves are directed. We demonstrate, both analytically and numerically, that interlayer directionality alone can reproduce all previously known diffusion regimes, including superdiffusion and the prime regime. More importantly, we unveil a new dynamical behavior—directionality-induced jamming—in which the asymmetric orientation of interlayer links suppresses diffusion and dynamically fragments the system into disconnected components, preventing convergence to equilibrium. We further show that this regime can arise not only in synthetic models but also in real-world topologies such as the London transport multiplex \cite{Domenico2014}. Our findings reveal that interlayer directionality constitutes a fundamental control parameter shaping diffusion in interconnected systems, opening new perspectives for the design, optimization, and robustness analysis of multilayer networks.

%\section{Formalism}

We now introduce the notation and framework used to describe diffusion dynamics on multiplex networks with directed interlayer couplings. Consider a two-layer multiplex of $N$ nodes per layer. The state vector representing the concentration on the $2N$ nodes at time $t$ is:
\begin{equation}
    \textbf{x}(t)=(\textbf{x}_1|\textbf{x}_2)^T
    \label{eq::StateVector}
\end{equation}
The diffusion dynamics in such a system are governed by
\begin{equation}
\frac{d\mathbf{x}}{dt}=-\mathcal{L}\mathbf{x},
\label{DiffeqLap}
\end{equation}
where $\mathcal{L}\in\mathbb{R}^{2N\times 2N}$ is the so-called \textit{supraLaplacian} matrix, defined for multiplex consisting of two layers as
\begin{equation}
\mathcal{L}=
\left(
\begin{array}{c|c}
D_1L_1+D_{12}I_N & -D_{12}I_N \\
\hline
-D_{21}I_N & D_2L_2+D_{21}I_N
\end{array}
\right).
\label{Lap}
\end{equation}
Here, $D_1$ and $D_2$ are the diffusion coefficients within each layer, $D_{12}$ and $D_{21}$ represent interlayer diffusion, $L_1$ and $L_2$ are the Laplacian matrices of the respective layers, and $I_N$ denotes the $N\times N$ identity matrix. 

We denote by $\{\Lambda_1 \leq \Lambda_2 \leq \cdots \leq \Lambda_{2N}\}$ the ordered set of eigenvalues of the supra-Laplacian $\mathcal{L}$, and by $\{\lambda_1^{(k)} \leq \lambda_2^{(k)} \leq \cdots \leq \lambda_N^{(k)}\}$ the eigenvalues of the Laplacian $L_k$ of the $k$-th layer. In the presence of directionality, eigenvalues are ordered according to their real parts. For a multiplex that forms a strongly connected component, the diffusion dynamics converge to the eigenvector associated with the zero eigenvalue, $\lambda_1 = 0$ (and thus $\Lambda_1 = 0$). This eigenvector corresponds to a uniform concentration distribution across all nodes (i.e., all 1s). The rate of convergence toward this steady state is determined by the real part of the smallest nonzero eigenvalue, $\lambda_2$ (or $\Lambda_2$ for the multiplex), which governs the slowest relaxation mode of the system.

%From this formulation, it follows that the system admits solutions of the form $\phi_i(t) = \phi_i(0)e^{-\Lambda_i t}$, where $\Lambda_i$ ($i=1,...,2N$) are the eigenvalues of $\mathcal{L}$, with associated eigenvectors $\phi_i$. If the network is connected, the convergence rate is determined by $\text{Re}(\Lambda_2)$, since $\Lambda_1=0$ due to $\mathcal{L}$ being a Laplacian matrix \cite{Newman2010}.

We now address how directionality can be incorporated into the interlayer coupling. Two distinct mechanisms are considered: \emph{topological directionality} and \emph{induced directionality}. In the first case, directionality is encoded explicitly in the interlayer topology: the presence of a link from node $n_i^{(k)}$ in layer $k$ to its counterpart $n_i^{(l)}$ in layer $l$ does not require the reciprocal link from $n_i^{(l)}$ to $n_i^{(k)}$. The supra-adjacency matrix of such a multiplex can therefore be written as
\begin{equation}
\widetilde{A} =
\begin{pmatrix}
A_1 & \Delta_1 \\
\Delta_2 & A_2
\end{pmatrix} \in \mathbb{R}^{2N \times 2N},
\label{supra}
\end{equation}
where $A_K$ denotes the adjacency matrix of layer $K \in \{1,2\}$ (assumed symmetric here), and $\Delta_K$ is a diagonal matrix whose elements $(\Delta_K)_{ii} \in \{0,1\}$ encode the existence of interlayer links. Consequently, in Eq.~(3), the identity matrices $I_N$ appearing in the interlayer blocks must be replaced by $\Delta_1$ and $\Delta_2$ to correctly account for directionality. In contrast, \emph{induced directionality} arises when interlayer links are topologically undirected but exhibit asymmetric diffusion coefficients. Retaining the structure of Eq.~(3) while imposing $D_{12} \neq D_{21}$ introduces an effective directionality in the coupling between layers, and thus in the overall diffusion dynamics.

In the following, we characterize the effects of interlayer directionality on diffusion dynamics, combining analytical derivations with numerical experiments. We first examine the case of induced directionality, where asymmetry arises solely from unequal interlayer diffusion coefficients ($D_{12} \neq D_{21}$), and subsequently address topological directionality, where the interlayer structure itself imposes directed connectivity. This separation allows us to isolate the distinct physical mechanisms by which interlayer orientation shapes the emergent dynamical regimes of multiplex systems.

%\section{Induced directionality}

%\section*{Induced Directionality}

We first consider the case where directionality is introduced through asymmetric interlayer diffusion coefficients. In this setting, the supra-Laplacian takes the form
\begin{equation}
\mathcal{L} =
\begin{pmatrix}
L_1 & 0 \\
0 & L_2
\end{pmatrix}
+ D_{12}
\begin{pmatrix}
I & -I \\
0 & 0
\end{pmatrix}
+ D_{21}
\begin{pmatrix}
0 & 0 \\
-I & I
\end{pmatrix}.
\label{eq:L_induced}
\end{equation}

Without loss of generality, we set $D_1 = D_2 = 1$ to simplify notation. Since our focus is on the effect of interlayer coupling, it is convenient to work in the basis $(\textbf{1}|\textbf{1})^T$ and $(-\textbf{1}|\textbf{1})^T$, corresponding to the eigenvectors of the coupling matrices. In this representation, the transformed supra-Laplacian reads
$$
\hat{\mathcal{L}} = U\mathcal{L}U^{-1} =
\frac{1}{\sqrt{2}}\left(
\begin{array}{cc}
1& 1 \\
-1 & 1
\end{array}
\right)\mathcal{L}\left(\begin{array}{cc}
1& -1 \\
1 & 1
\end{array}
\right)\frac{1}{\sqrt{2}} =
$$\begin{equation}
=\left(
\begin{array}{cc}
L_+ & L_-+(D_{21}-D_{12})I \\
L_- & L_++(D_{12}+D_{21})I
\end{array}
\right),
\label{eq:Lhat}
\end{equation}
where $L_{+} = \tfrac{1}{2}(L_1 + L_2)$ and $L_{-} = \tfrac{1}{2}(L_2 - L_1)$, with eigenvalues $\lambda_j^{(+)}$ and $\lambda_j^{(-)}$, respectively. Applying the Schur complement \cite{Cottle1974} to $|\hat{\mathcal{L}}-\Lambda I|=0$ yields

%\resizebox{\linewidth}{!}{
%\begin{equation}
%\begin{aligned}
%& \\
%& |L_++(D_{12}+D_{21})I-\Lambda I| \cdot \\
%& |L_+-\Lambda I-(L_-+(D_{21}-D_{12})I)(L_++(D_{12}+D_{21})I-\Lambda I)^{-1}L_-| =0 \\
%&
%\end{aligned}
%\label{EigEqHeat}
%\end{equation}
%}

\begin{widetext}
\begin{equation}
\label{EigEqHeat}
\left|L_+ + (D_{12}+D_{21})I - \Lambda I\right|
\,\left|L_+ - \Lambda I
- \big(L_- + (D_{21}-D_{12})I\big)
\big(L_+ + (D_{12}+D_{21})I - \Lambda I\big)^{-1}
L_- \right| = 0.
\end{equation}
\end{widetext}

From Eq.~\eqref{EigEqHeat}, two asymptotic regimes follow:

\begin{enumerate}
    \item For $D_{12} \gg D_{21}, \, D_{12} \gg 1$ (or more strictly, $D_{12}\gg N$, see the Supplementary Materials for more details), the equation reduces to $|L_++D_{12}I-\Lambda I|\cdot |L_2-\Lambda I|=0$ and the spectrum separates into eigenvalues scaling with $D_{12}$ and those converging to the eigenvalues of $L_2$. The smallest nonzero eigenvalue $\Lambda_2$ thus approaches $\lambda_2^{(2)}$, and the slower layer (layer 2) controls the overall convergence rate.

    \item Conversely, for $D_{21} \gg D_{12}, \, D_{21} \gg 1$ (again see Supplementary Material for the details of the more stricter condition $D_{21}\gg N$), the equation becomes $|L_++D_{21}I-\Lambda I|\cdot |L_1-\Lambda I|=0$ and the situation reverses: $\Lambda_2 \rightarrow \lambda_2^{(1)}$, and diffusion is dictated by layer~1.
\end{enumerate}

\begin{figure}[!t]
\centering
\includegraphics[width=0.99\columnwidth]{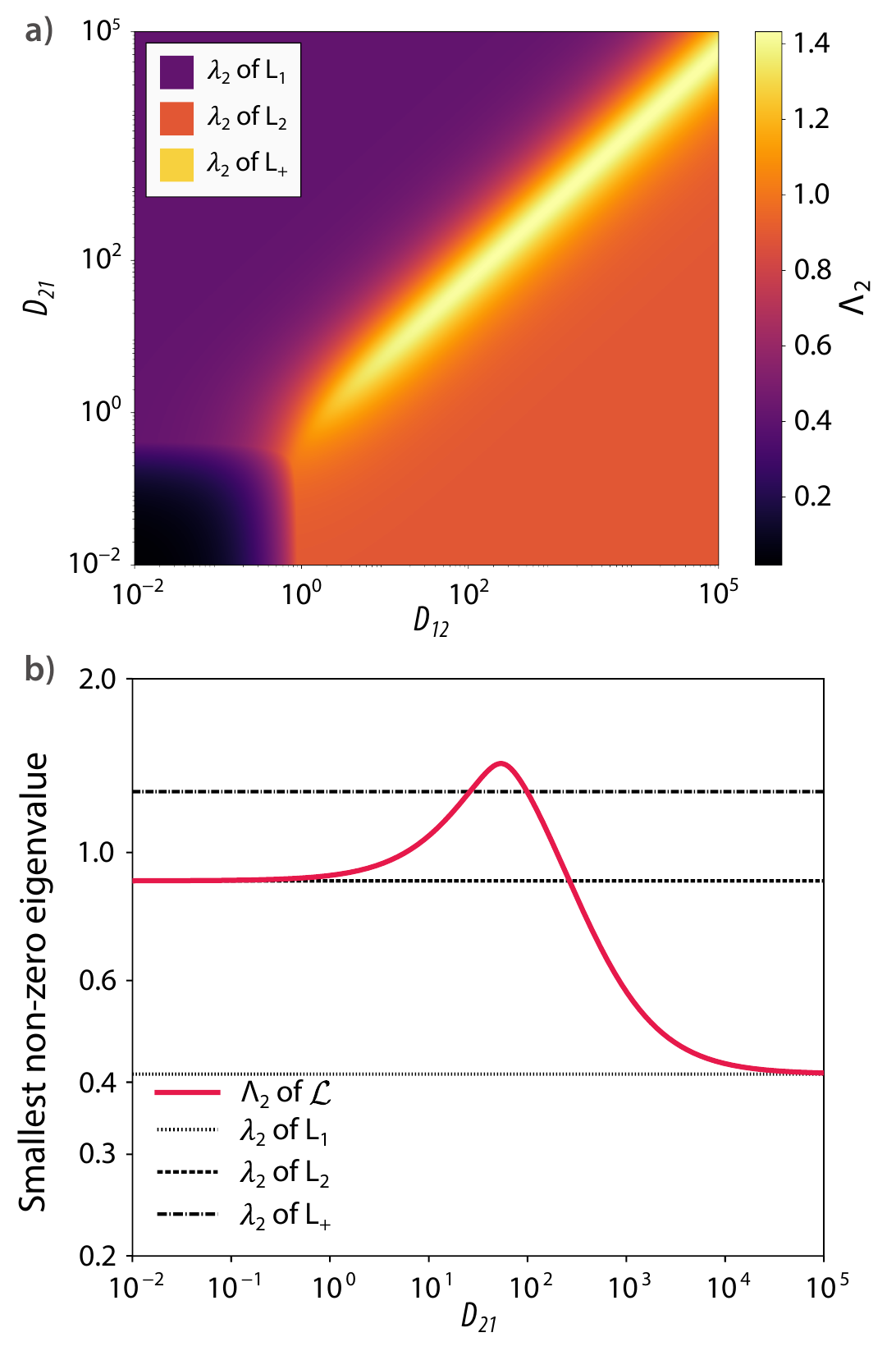}
\caption{\raggedright  \textbf{Diffusion Dynamics in multiplex networks with induced interlayer directionality.} a) Heat map of the smallest non-zero eigenvalue of $\mathcal{L}$, $\Lambda_2$, as a function of the coupling parameters $D_{12}$ and $D_{21}$. b) Evolution of $\Lambda_2$ as function of $D_{21}$ when $D_{12}=100.0$. The network is given in the Supplementary Materials.}
\label{Fig1}
\end{figure}

The existence of these two limiting cases and because the eigenvalues vary continuously with $D_{12}$ and $D_{21}$ \cite{horn2013matrix} directly imply that the system must transition smoothly between such regimes, passing through the symmetric case $D_{12} = D_{21}$ studied in Ref.~\cite{Gomez2013}, where $\Lambda_2 = \lambda_2^{(+)}$. Moreover, if $\lambda_2^{(+)}>\max{(\lambda_2^{(1)},\lambda_2^{(2)})}$, $\Lambda_2$ would exhibit non-monotonic behavior, which ensures the existence of the prime regime as well. Consequently, induced directionality can reproduce both \emph{superdiffusion} and the \emph{prime regime} reported in previous works~\cite{Tejedor2018, Wang2021}. The numerical validation of these regimes for the network described in the Supplementary Material is shown in Fig.~\ref{Fig1}.

We now turn to the case where directionality is encoded explicitly in the interlayer topology. As defined in Eq.~(\ref{supra}), the off-diagonal blocks of the supra-adjacency matrix are non-symmetric, reflecting that the interlayer connection from layer $k$ to layer $l$ does not necessarily imply its reciprocal. Assuming $D_{12} = D_{21} = D_X$ and $D_1 = D_2 = 1$ for simplicity, the supra-Laplacian reads
\begin{equation}
\mathcal{L} =
\begin{pmatrix}
L_1 & 0 \\
0 & L_2
\end{pmatrix}
+ D_X
\begin{pmatrix}
\Delta_1 & -\Delta_1 \\
-\Delta_2 & \Delta_2
\end{pmatrix}.
\label{eq:L_topological}
\end{equation}

For $D_X = 0$, the eigenvalue spectrum of $\mathcal{L}$ is simply the union of the spectra of $L_1$ and $L_2$. To understand how coupling modifies this spectrum, we analyze the weak and strong coupling limits.

\paragraph{Weak coupling ($D_X \to 0$)}. In this limit, the interlayer term acts as a perturbation to the block-diagonal matrix $\mathcal{L}_0 = \mathrm{diag}(L_1, L_2)$. Because both $L_1$ and $L_2$ are Laplacians, $\mathcal{L}_0$ possesses a degenerate zero eigenvalue of minimum multiplicity $g_0 = 2$. The first-order correction to the degenerate eigenvalues, $\Lambda^g(1)$, can be expressed as: 
\begin{equation}
\Lambda^g(1) = \Lambda^g(0)+\mu_g D_X + o(D_X)
\label{eq:jamming2}
\end{equation}
where $g=\{1,...,g_0\}$, $\Lambda^g(0)$ is the unperturbed eigenvalue (here, $\Lambda^g(0)=0$), and $\mu_g$ are the eigenvalues of the perturbation matrix $W\in\mathbb{C}^{g_0\times g_0}$ whose elements are given by $W_{kl}=v^T_k\mathcal{D}v_l$. Since the eigenvectors $v_i$ that span the kernel of $\mathcal{L}_0$ are  $(\textbf{1}|\textbf{0})^T$ and $(\textbf{0}|\textbf{1})^T$, the explicit construction of W yields
\begin{equation}
    W=\frac{1}{N}\left(\begin{array}{cc}
     \sum_{i=1}^{N}(\Delta_1)_{ii} & -\sum_{i=1}^{N}(\Delta_1)_{ii}\\
    -\sum_{i=1}^{N}(\Delta_2)_{ii} & \sum_{i=1}^{N}(\Delta_2)_{ii}
    \end{array}\right).
\end{equation}
whose eigenvalues are $\mu_1 = 0$ and $\mu_2 = N^{-1}\sum_i [(\Delta_1)_{ii} + (\Delta_2)_{ii}]$. Consequently, the smallest nonzero eigenvalue of $\mathcal{L}$ grows linearly with the coupling as
\begin{equation}
\Lambda_2 = \mu_2 D_X + o(D_X).
\end{equation}
indicating that for weak coupling, the relaxation rate increases linearly with $D_X$.

\paragraph{Strong coupling ($D_X \to \infty$)}. In the opposite limit, we analyze the characteristic polynomial $|\widehat{\mathcal{L}} - \Lambda I| = 0$ in the transformed basis used in Eq.~(\ref{eq:Lhat}). This yields
$$
\left|\begin{array}{cc}
L_+-\Lambda I& L_-+D_X\Delta_- \\
L_- & L_++D_X\Delta_+-\Lambda I
\end{array}
\right|=
$$$|L_++D_X\Delta_+-\Lambda I|\cdot$
\begin{equation}
|L_+-\Lambda I-(L_-+D_X\Delta_-)(L_++D_X\Delta_+-\Lambda I)^{-1}L_-|=0
\label{eq:det3}
\end{equation}
where $\Delta_+=(\Delta_1+\Delta_2)$ and $\Delta_-=(\Delta_2-\Delta_1)$. The first term, $|L_++D_X\Delta_+-\Lambda I|$, produces eigenvalues that diverge linearly with $D_X$, with slopes set by the diagonal entries of $\Delta_+$, which can be 1 or 2 (Note that the cases with zero value for $\Delta_+$  are not considered so $\Delta_+^{-1}$ is well defined). The second factor, $|L_+-\Lambda I-(L_-+D_X\Delta_-)(L_++D_X\Delta_+-\Lambda I)^{-1}L_-|$, is more intricate. In the asymptotic limit $D_X\to\infty$, assuming $\Delta_+$ is invertible, this term converges to $|L_+-\Delta_-\Delta_+^{-1}L_--\Lambda I|$. This indicates that the second set of eigenvalues, $\lambda_i^{(\pm)}$, with $i=1,...,N$, converges to the spectrum of the matrix $L_+-\Delta_-\Delta_+^{-1}L_-$, which is Hermitian, hence, its eigenvalues are real and finite. Additionally, it can be proven that $0 \leq \lambda^{(\pm)}_i \leq \lambda_{\text{max}}(L_+)-\lambda_{N-i+1}(\Delta_-\Delta_+^{-1}L_-)$ (see Supplementary Materials for the proof).

A key consequence is that the zero eigenvalue can recover a degeneracy of two or higher at arbitrarily large values of $D_X$, implying the emergence of a new regime in which diffusion effectively stalls. In this \emph{directionality-induced jamming} regime, the system becomes dynamically partitioned into disconnected components, preventing convergence to a uniform steady state. Admittedly, the convergence rate to the finite eigenvalues can be determined through perturbation analysis. By rewriting $\mathcal{L}$ as follows, with $\epsilon=1/D_X$:
\begin{equation}
\mathcal{L}= D_X(\mathcal{D}+\epsilon\mathcal{L}_0)=D_X\left[\left(
\begin{array}{cc}
\Delta_1 & -\Delta_1 \\
-\Delta_2 & \Delta_2
\end{array}
\right)+ \epsilon\left(
\begin{array}{cc}
L_1& 0 \\
0 & L_2
\end{array}
\right)\right]
\end{equation} 
we can treat $\epsilon\mathcal{L}_0$ as a perturbation of $\mathcal{D}$. In this case, the first-order expansion for the eigenvalue $\Lambda_2$, (whose unperturbed value $\Lambda_2(0)=0$) yields
\begin{equation}
\Lambda_2 = D_X\left[\Lambda_2(0)+\epsilon\mu_2 + \mathcal{O}(\epsilon^2)\right] =  \mathcal{O}(1/D_X).
\end{equation} 
That is, $\Lambda_2$ decreases at a rate of at least $D_X^{-1}$.

\begin{figure*}[ht]
\centering
\hspace{-1cm} % Ajusta este valor para desplazar la imagen a la izquierda
\includegraphics[width=0.9\textwidth, height=3.5in]{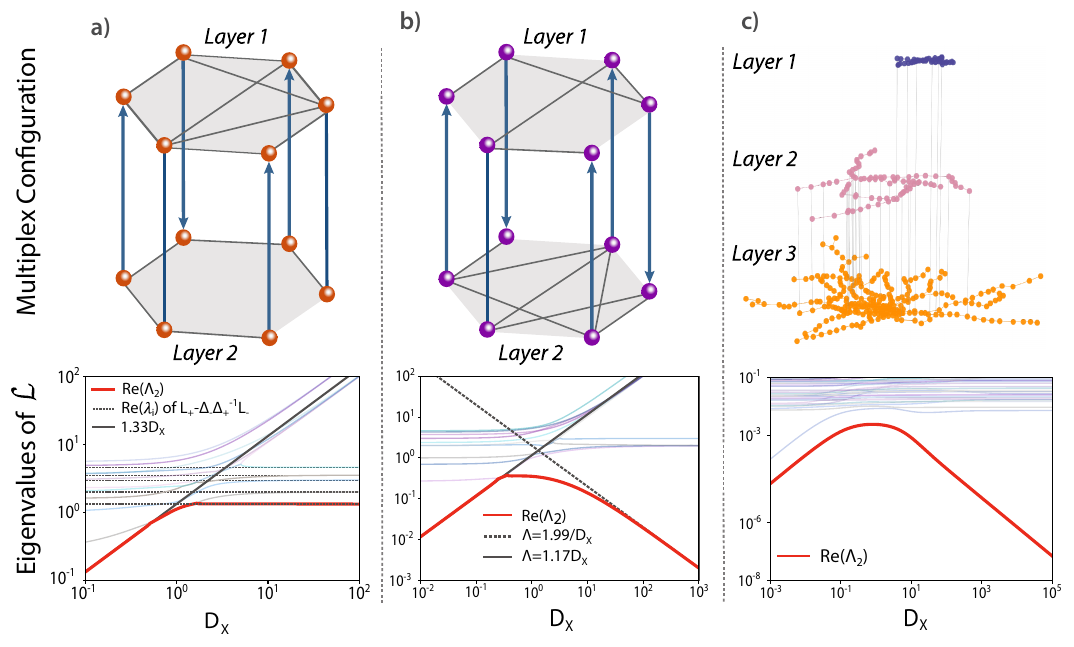}
\caption{\raggedright  \textbf{Diffusion Dynamics in multiplex networks with topological interlayer directionality}. Evolution of the eigenvalues of $\mathcal{L}$ versus the coupling parameter $D_X$ for different network configurations: a) and b) are synthetic networks of $N=6$ nodes per layer. c) represents the London transport network \cite{Domenico2014}, where the nodes are the stations and the links are the connections between them. Layer 1 represents the DLR line, layer 2 the overground, and layer 3 the underground. It is assumed that $D_K=1.0$ for all layers of any configuration.}
\label{Fig:Toy_plus_London}
\end{figure*}

Figure~\ref{Fig:Toy_plus_London} illustrates these effects for the synthetic topology depicted in Fig. \ref{Fig:Toy_plus_London}a,b, here used as an illustrative example. Furthermore, to explore the generality and potential practical relevance of these findings, we investigate whether it is possible to induce the jamming regime in a real-world system such as the London transportation system, which has been previously modeled as a multiplex network \cite{Domenico2014}. Since the network structure is predetermined, we examined the emergence of directionality-induced jamming by modifying the weights of the interlinks. In particular, we considered the problem as an optimization problem via a simulated annealing algorithm, wherein we seek interlayer link weights that maximize the likelihood of obtaining a jamming regime. We show in Fig. \ref{Fig:Toy_plus_London}c how it is possible to find distributions of interlayer link weights that lead to a jamming regime in real-world topologies (for more details, see Supplementary Materials).

In summary, we have shown that directionality in the coupling between layers of a multiplex network profoundly alters diffusion dynamics. Through analytical derivations and numerical experiments, we demonstrated that interlayer directionality alone—without any asymmetry within layers—can reproduce both the superdiffusion and prime regimes previously reported for directed multiplexes. More importantly, we uncovered a new dynamical phase, the \emph{directionality-induced jamming} regime, in which strong asymmetric couplings hinder diffusion and dynamically fragment the system into effectively disconnected components. This jamming transition emerges from purely structural asymmetry and is characterized by the vanishing of the smallest nonzero eigenvalue of the supra-Laplacian as $\Lambda_2 \sim D_X^{-1}$. Moreover, the disconnected components might map quite closely to individual layers, but generally involve nodes of different layers, making it difficult to predict a priori the structural or dynamical conditions (e.g., layer topology) under which jamming will occur. Addressing this challenge remains a key open problem that we expect will spur further research.

Finally, our findings identify interlayer directionality as a fundamental control parameter governing transport, relaxation, and connectivity in multiplex systems. Beyond their theoretical relevance, they have direct implications for the design and regulation of interconnected infrastructures $-$such as transportation and communication networks$-$, where directionality can be engineered or exploited to either enhance or suppress diffusion. We expect the concept of directionality-induced jamming to stimulate further exploration of how orientation and asymmetry shape collective dynamics (e.g., information or rumor spreading) in complex multilayer systems.

\paragraph{Acknowledgements}. Y.M. and A.T. were partially supported by the Government of Aragón, Spain, and ``ERDF A way of making Europe'' through grant E36-23R (FENOL), and by Ministerio de Ciencia, Innovación y Universidades, Agencia Española de Investigación (MICIU/AEI/ 10.13039/501100011033) Grant No. PID2023-149409NB-I00. A.T. was also partially supported by NSF Grants EAR-2342937, RISE-2425932, and RISE-2425748. 

%\bibliographystyle{apsrev4-1} % Tell bibtex which bibliography style to use
%\bibliography{Bib_Multiplex}

%

\clearpage
\appendix
\onecolumngrid
\begin{center}
  \textbf{\large Supplemental Material for ``Directionality-Induced Jamming in Multiplex Networks''}
\end{center}

\section{Topology of the Toy multiplex network to simulate induced interlayer directionality. }
\begin{figure}[h]
    \centering\includegraphics[width=0.2\linewidth]{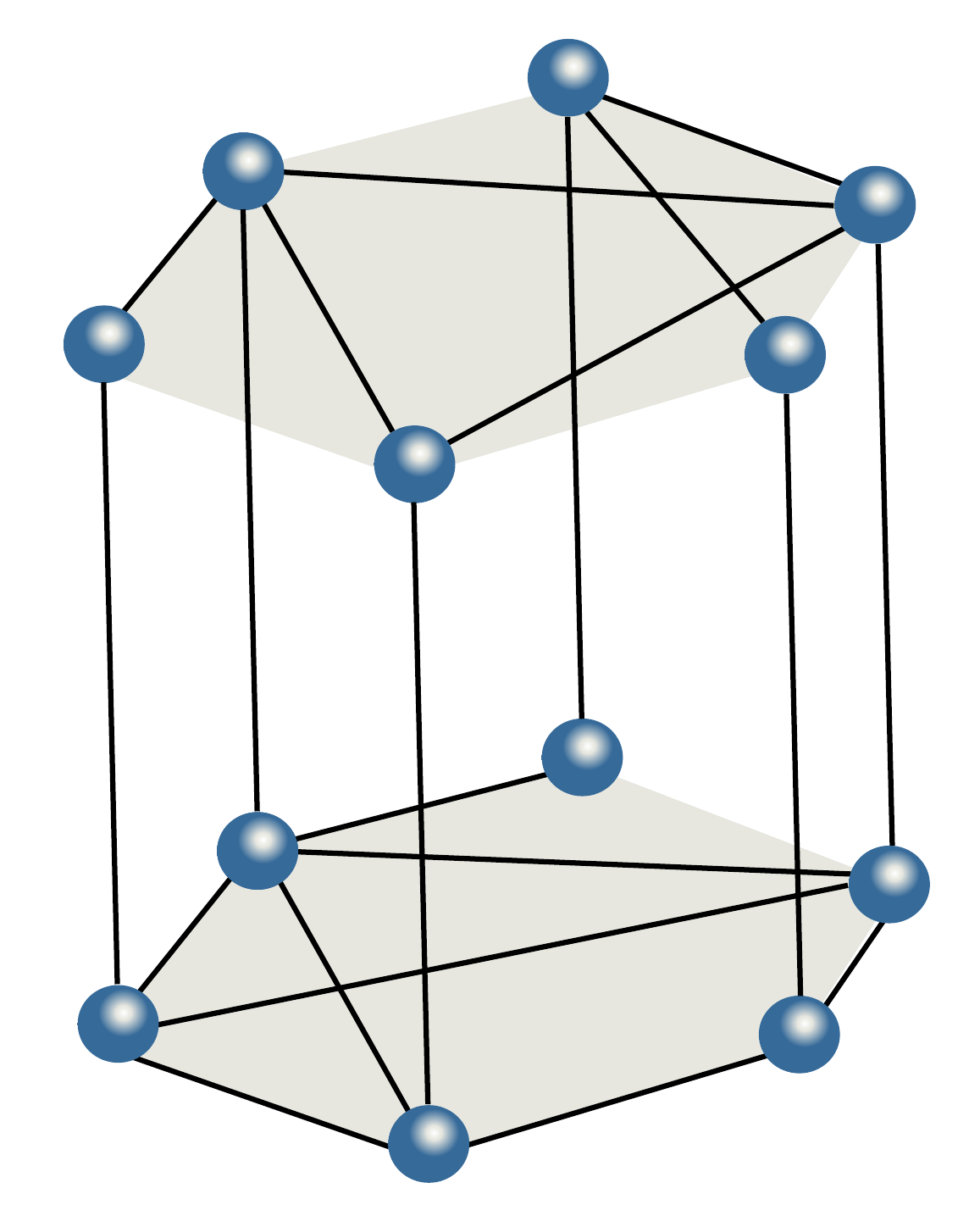}
    \caption{Topology of the multiplex network used for the results presented in Fig. 1 of the main manuscript.}
    \label{fig:Fig1SM}
\end{figure}

The multiplex network topology used to generate the results shown in Figure 1 of the main manuscript is shown in Fig. \ref{fig:Fig1SM}. This topology can be generated using the ``Code.ipynb" (the instructions are provided there).

\section{Approximation for Dominant Coupling $D_{12}$ in an induced directionality scenario}

Let us study the limit where $D_{12} \gg D_{21}$ and $D_{12} \gg N$ for the expression

\begin{equation}
    |L_++(D_{12}+D_{21})I-\Lambda I|\cdot |L_+-\Lambda I-(L_-+(D_{21}-D_{12})I)(L_++(D_{12}+D_{21})I-\Lambda I)^{-1}L_-|=0
\end{equation}

Imposing first that $D_{12} \gg D_{21}$, the expression simplifies to

\begin{equation}
   |L_++D_{12}I-\Lambda I|\cdot |L_+-\Lambda I-(L_--D_{12}I)(L_++D_{12}I-\Lambda I)^{-1}L_-|=0 
\end{equation}

Now, given that $D_{12} \gg N$, the term $D_{12}I$ dominates the sums (i.e., $L_-$, $L_+$, and $\Lambda I$ are asymptotically negligible compared to $D_{12}I$ because the maximum value of any component of these laplacians is $N$), which leads to

\begin{equation}
    |L_++D_{12}I-\Lambda I|\cdot |L_+-\Lambda I+D_{12}I(D_{12}I)^{-1}L_-|=0
\end{equation}

Simplifying, we obtain

\begin{equation}
    |L_++D_{12}I-\Lambda I|\cdot |L_+-\Lambda I+L_-|=0
\end{equation}

Since $L_+ = \tfrac{1}{2}(L_1 + L_2)$ and $L_- = \tfrac{1}{2}(L_2 - L_1)$, we finally arrive at

\begin{equation}
    |L_++D_{12}I-\Lambda I|\cdot |L_2-\Lambda I|=0
\end{equation}

\section{Proof of condition over $\lambda_i^{(\pm)}$}
We will now prove the following condition:
\begin{equation}
0\leq\lambda_i^{(\pm)}\leq\lambda_\text{max}(L_+)-\lambda_{N-i+1}(\Delta_-\Delta_+^{-1}L_-).
\end{equation}\\

Let us define the matrix $M=L_+-\Delta_-\Delta_+^{-1}L_-$. In this specific case, $L_+$ and $L_-$ are Hermitian matrices, while $\Delta_-$ and $\Delta_+^{-1}$ are diagonal matrices. As they have real components, the result $M$ must also be a Hermitian matrix. This property ensures that all of its eigenvalues are real and, therefore, can be ordered:

\begin{equation}
\label{eq:orden} \lambda_{\text{min}}=\lambda_1\leq\cdots\leq\lambda_N=\lambda_{\text{max}},
\end{equation}
where $N$ is the size of the eigenvalue set. To specify the matrix to which an eigenvalue belongs, we will use the notation $\lambda_i(A)$ for the eigenvalues of matrix A.

To prove the condition, we will use two established results for Hermitian matrices.

\begin{enumerate}
    \item \textbf{Eigenvalues of} $\mathbf{-A}$: 
    
    The first result states that if we have a Hermitian matrix $A$ with eigenvalues ordered as in Expression \ref{eq:orden}, then the ordered eigenvalues of $-A$ are expressed as:
    \begin{equation}
    -\lambda_N(A)\leq\cdots\leq-\lambda_1(A),
    \end{equation}
    which implies the relationship $\lambda_i(-A)=-\lambda_{N-i+1}(A)$ \cite{horn2013matrix}.
    
    \item \textbf{Weyl's Theorem Corollary}:

    The second result is a corollary of Weyl's Theorem, which states that for any two Hermitian matrices $A$ and $B$, the following inequality holds:    
    \begin{equation}
    \label{eq:Weyl}
    \lambda_i(A)+\lambda_1(B)\leq\lambda_i(A+B)\leq\lambda_i(A)+\lambda_N(B), \quad i=1,..., N
    \end{equation}
\end{enumerate}
    
By setting $A=-\Delta_-\Delta_+^{-1}L_-$ and $B=L_+$, we can apply Expression \ref{eq:Weyl} to affirm:

\begin{equation}
\lambda_i(-\Delta_-\Delta_+^{-1}L_-)+\lambda_1(L_+)\leq\lambda_i(M)\leq\lambda_i(-\Delta_-\Delta_+^{-1}L_-)+\lambda_N(L_+).
\end{equation}

Next, we must consider two important points. First, by construction, $\lambda_1(L_+)=0$ and $\lambda_N(L_+)=\lambda_{\text{max}}(L_+)$. Second, we can use the relationship $\lambda_i(-A)=-\lambda_{N-i+1}(A)$. Applying these points, we find:

\begin{equation}
\label{eq:46}
-\lambda_{N-i+1}(\Delta_-\Delta_+^{-1}L_-)\leq\lambda_i(M)\leq-\lambda_{N-i+1}(\Delta_-\Delta_+^{-1}L_-)+\lambda_\text{max}(L_+).
\end{equation}

For the system to be stable, we know that $\lambda_i(M)\geq 0$. This requires that the upper bound must also be non-negative, meaning $-\lambda_{N-i+1}(\Delta_-\Delta_+^{-1}L_-)+\lambda_\text{max}(L_+)\geq0$. This implies $-\lambda_\text{max}(L_+)\leq-\lambda_{N-i+1}(\Delta_-\Delta_+^{-1}L_-)$. Given this, we can rewrite the Expression \ref{eq:46} as:

\begin{equation}
-\lambda_\text{max}(L_+)\leq\lambda_i(M)\leq-\lambda_{N-i+1}(\Delta_-\Delta_+^{-1}L_-)+\lambda_\text{max}(L_+).
\end{equation}

The advantage of this form is that we know $\lambda_\text{max}(L_+)\geq0$. Leveraging the system's stability again, we can adjust the lower bound to 0, which leads to our final expression:

\begin{equation}
0\leq\lambda_i(M)\leq\lambda_\text{max}(L_+)-\lambda_{N-i+1}(\Delta_-\Delta_+^{-1}L_-).
\end{equation}

This derivation makes it clear that for all $i$, the eigenvalue 0 is a permissible value, which allows for its multiplicity to be greater than 1.

\section{Optimization of interlayer link directionality in the London multiplex to induce a jamming regime}

The goal of this optimization is to find configurations for a multiplex with a real topology (like the London traffic network) that exhibit a Jamming Regime. To achieve this, the aim was to find a configuration of weights, between 0 and 1, that would lead to this objective.

For this purpose, \textit{simulated annealing} was used as the optimization method, defining the energy function as:
% \cite{kirkpatrick1983optimization} 

$$E = \left|\frac{\text{Re}\left[\Lambda_2(D_X\to\infty)\right]}{\text{Re}\left[\Lambda_2(D_X\to\infty)\right]-\max(\text{Re}\left[\Lambda_2\right])}\right|$$

This energy function reaches its minimum value when $\text{Re}\left[\Lambda_2(D_X\to\infty)\right]\to 0$, while also favoring the largest possible difference between this value and the maximum value of $\Lambda_2$. These two conditions are the ones satisfied in the jamming regime.

Additionally, during the optimization, it was required that the network remain strongly connected at all times. This prevents the found configuration from blocking diffusion by disconnecting the network into several components.

\begin{figure}[h]
    \centering
    \includegraphics[width=1.0\linewidth]{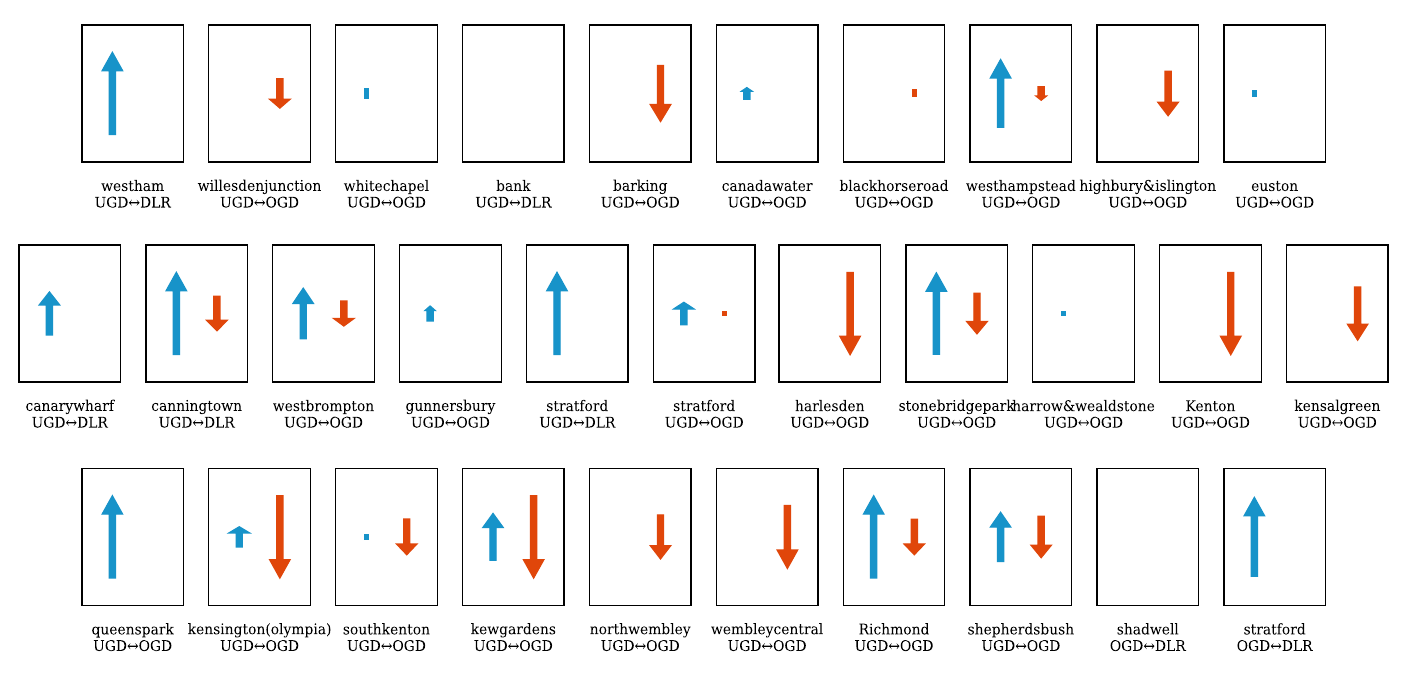}
    \caption{Representation of the interlink values. DLR represents the Docklands Light Railway (Layer 1), OGD the overground (Layer 2), and UGD the underground (Layer 3). The size of the arrow represents its value (with a maximum of 1 and a minimum of 0), and its color (along with its direction) represents the direction of the link: blue represents the normal direction (for example, westham UGD $\rightarrow$ DLR in the first case) and red represents the reverse direction (westham DLR $\rightarrow$ UGD in the same example). The numerical values are given in the file "interlinks\_values.txt".}
    \label{fig:placeholder}
\end{figure}

With all this in mind, the interlayer weights are shown in Fig.\ref{fig:placeholder}.

We provide the modified adjacency matrix (with the interlinks already weighted) in the file  ``modified\_matrix\_jamming.csv".

If the unmodified matrix is required, it can be retrieved from ``Code.ipynb".

%\bibliography{Bib_Multiplex}

\end{document}